\documentclass[aps,prl,twocolumn,superscriptaddress,longbibliography,footinbib,nobibnotes]{revtex4-2}  
\usepackage{graphicx} 
\usepackage{float} 
\usepackage{dcolumn} 
\usepackage{bm,color}
\usepackage{amssymb}
\usepackage{amsmath}
\usepackage{extarrows}
\usepackage[colorlinks=true,linkcolor=blue,urlcolor=blue,citecolor=blue]{hyperref}

\usepackage[normalem]{ulem} 
\definecolor{mgreen}{RGB}{1,123,0}

\definecolor{felixgreen}{rgb}{0,0.75,0}

\definecolor{Nathanblue}{rgb}{0.,0.24,0.51}

\newcommand{\blue}{\color{Nathanblue}}

\begin{document}
\title{{\blue Anyon-Impurity Bound States in Quantum-Engineered Fractional Chern Insulators}}

\author{Botao Wang} 
\thanks{These authors contributed equally to this work.}
\affiliation{International Solvay Institutes, B-1050 Brussels, Belgium}
\affiliation{Center for Nonlinear Phenomena and Complex Systems,
Université Libre de Bruxelles, CP 231, Campus Plaine, B-1050 Brussels, Belgium}

\author{Amit Vashisht}
\thanks{These authors contributed equally to this work.}
\affiliation{International Solvay Institutes, B-1050 Brussels, Belgium}
\affiliation{Center for Nonlinear Phenomena and Complex Systems,
Université Libre de Bruxelles, CP 231, Campus Plaine, B-1050 Brussels, Belgium}

\author{Felix A. Palm}
\affiliation{International Solvay Institutes, B-1050 Brussels, Belgium}
\affiliation{Center for Nonlinear Phenomena and Complex Systems,
Université Libre de Bruxelles, CP 231, Campus Plaine, B-1050 Brussels, Belgium}

\author{Fabian~Grusdt}
\affiliation{Department of Physics and Arnold Sommerfeld Center for Theoretical Physics (ASC), Ludwig-Maximilians-Universität München, Theresienstr. 37, D-80333 München, Germany}
\affiliation{Munich Center for Quantum Science and Technology (MCQST), Schellingstr. 4, D-80799 München, Germany}

\author{Laurens Vanderstraeten}
\affiliation{Center for Nonlinear Phenomena and Complex Systems,
Université Libre de Bruxelles, CP 231, Campus Plaine, B-1050 Brussels, Belgium}

\author{Nathan Goldman}
\email{nathan.goldman@lkb.ens.fr}
\affiliation{International Solvay Institutes, B-1050 Brussels, Belgium}
\affiliation{Center for Nonlinear Phenomena and Complex Systems,
Université Libre de Bruxelles, CP 231, Campus Plaine, B-1050 Brussels, Belgium}
\affiliation{Laboratoire Kastler Brossel, Collège de France, CNRS, ENS-Université PSL,
Sorbonne Université, 11 Place Marcelin Berthelot, 75005 Paris, France}

\date{\today}

\begin{abstract}
	 Mobile impurities provide a powerful means of probing correlated and topological quantum matter, through their dressing by the surrounding medium and the practical probes granting access to the resulting composite object. Motivated by the recent observation of anyon-impurity composites in the solid state, as well as recent realizations of Laughlin-type states in engineered lattice systems, we investigate the formation of a bound state between a mobile impurity and a single pinned quasihole in the interacting Harper–Hofstadter model deep in the fractional Chern insulator regime. Combining analytical arguments with large-scale numerical simulations, we characterize the structure, energetics, and stability of hybrid anyon-impurity bound states, and show that their binding energy provides direct access to the fractional charge of the quasihole under conditions that we identify. We further demonstrate that the composite object can be coherently transported by externally steering the quasihole pinning potential. Our results establish a realistic pathway for controlled anyon-impurity manipulation in quantum-engineered platforms, enabling experimentally feasible protocols for braiding.
\end{abstract}

\maketitle

\paragraph{Introduction---}  Mobile impurities have long served as powerful probes of quantum many‑body systems, from classic polaron physics to modern cold‑atom realizations where interacting mixtures can be controlled and analyzed with high precision~\cite{grusdt2025impurities,2026Massignan}. Particularly intriguing is the study of impurities interacting with topological quantum matter, where the impurity can inherit geometric or topological properties from the medium and couple to its low-energy excitations~\cite{excitons_QH_dressing_1992,Anyon_excitons1993,2014Zhang,Grusdt2016,2019Guardian,Grass_2020,2020Heras,2021Pimenov,2021Baldelli,Vashisht2025,Anton_impurity_Majorana,sadovnikov2026localizationpatternmobileimpurity,Mostaan2026,Wagner2026,2026Lu}.

An important direction concerns the binding of impurities to fractionalized excitations in fractional quantum Hall (FQH) systems~\cite{Anyon_excitons1993,2014Zhang,Grusdt2016,Grass_2020,2020Heras,2021Baldelli,2025Huang,2026Apalkov,Mostaan2026,Wagner2026,2026Lu}. Binding an impurity to a quasihole has been proposed as a means to measure the many‑body Chern number~\cite{Grusdt2016}, fractional charges and statistics~\cite{2020Heras,Grass_2020,Mostaan2026} and to manipulate anyonic excitations~\cite{Palm2026}. This prospect is especially compelling because quasiholes are anyons whose controlled motion underlies proposals for topological quantum information processing~\cite{2003Kitaev,2008Nayak}. Recent theoretical and experimental works have reinforced this vision: `anyon‑trions' -- bound states of quasiholes with localized excitons -- were predicted in semiconductor heterostructures~\cite{Mostaan2026,Wagner2026,2026Lu} and subsequently observed in twisted MoTe$_{\rm 2}$ in the fractional Chern insulating (FCI) regime~\cite{2026Li}. Local spectroscopy of anyon-impurity bound states has also been reported recently in monolayer graphene~\cite{2026Park}. However, whether such anyon-impurity composites could be created and manipulated in quantum-engineered settings, such as cold-atom~\cite{2019Cooper} or photonic~\cite{2019Ozawa_photon} platforms, remains an outstanding challenge.

In this Letter, we investigate the binding of a charge-neutral mobile impurity to a single pinned quasihole in the  interacting Harper-Hofstadter model deep in the FCI regime. This setting is ideally suited to cold-atom experiments~\cite{2008Bloch,2019Cooper,2023Leonard}, where quasiholes can be created and localized with high fidelity using programmable pinning potentials in a quantum‑gas microscope~\cite{2009Bakr,2010Sherson,2016Kuhr,2024Braun,2015Liu,Raciunas2018,Macaluso2020,Wang2022,Palm2024}, and where a mobile impurity can be realized as a single atom of a different species with tunable  impurity-host interactions~\cite{grusdt2025impurities,2026Massignan,2017Meinert,2025Dhar}. Combining analytical insight with large-scale numerical methods, we characterize the structure and stability of this hybrid impurity–anyon bound state and demonstrate controlled motion of the composite object via external manipulation of the pinning potential. Our results establish a realistic pathway toward precise anyon-impurity manipulation in cold‑atom platforms, opening the door to experimentally feasible braiding protocols in the bulk~\cite{Palm2026}.

\paragraph{Model---} We consider a single mobile `impurity' particle interacting with an ensemble of `host' particles forming a quantum Hall state in the interacting Harper-Hofstadter model. The total system is described by the Hamiltonian
\begin{equation}
	\text{\ensuremath{\hat{H}=}}\hat{H}_{{\rm host}}+\hat{H}_{{\rm imp}}+\hat{H}_{{\rm U}}.
	\label{H}
\end{equation}
Here, $\hat{H}_{{\rm host}}$ describes host particles moving in a two dimensional square lattice penetrated by a uniform magnetic flux,
\begin{align}
	\hat{H}_{{\rm host}}= & -J{\displaystyle \sum_{x,y}}(e^{i\phi y}\hat{b}_{x+1,y}^{\dagger}\hat{b}_{x,y}+\hat{b}_{x,y+1}^{\dagger}\hat{b}_{x,y}+{\rm H.c.}) \nonumber \\
	&+\hat{H}_{{\rm int}}^{{\rm B/F}}+V_{h}{\displaystyle \sum_{x,y\in\mathcal{P}}}\hat{n}_{x,y},
    \label{Hhost}
\end{align}
where $J$ is the hopping amplitude of the host particles and $\phi=2\pi\alpha$ is the magnetic flux per plaquette. The operators $\hat{b}_{x,y}$ ($\hat{b}^\dagger_{x,y}$) annihilate (create) a boson (B) or fermion (F) at lattice site $(x,y)$ with $x$ and $y$ being lattice indices along the horizontal and vertical directions, respectively. The number operators read $\hat{n}_{x,y}=\hat{b}^\dagger_{x,y}\hat{b}_{x,y}$.
In the fermionic case, we consider nearest-neighbor interactions $\hat{H}_{{\rm int}}^{{\rm F}}=V_{{\rm NN}} \sum_{x,y}(\hat{n}_{x+1,y}\hat{n}_{x,y}+\hat{n}_{x,y+1}\hat{n}_{x,y})$ with $V_{\rm NN}$ the interaction strength.  Both integer Chern insulator (CI) and FCI ground states can then be stabilized, depending on the filling factor and interaction strength~\cite{Pauw2026}.
To target a bosonic Laughlin-type ground state, we consider hard-core bosons: $\hat{H}_{{\rm int}}^{{\rm B}}=\frac{U_{h}}{2} \sum_{x,y}\hat{n}_{x,y}(\hat{n}_{x,y}-1)$ with  $U_h\rightarrow\infty$~\cite{Soerensen2005,2006Palmer,Hafezi2007,2017Gerster,Motruk2017,2018Dong,2019Rosson,Macaluso2020,2020Motruk,2020Repellin}. 
We use $V_h$ to denote an on-site pinning potential applied in a local region $\mathcal{P}$, for example, the central plaquette of the system. Such a potential is used to create localized integer and fractional charges in the CI and FCI states, respectively~\cite{2015Liu,Raciunas2018,Macaluso2020,Wang2022,Palm2024}.

The motion of a single charge-neutral impurity is described by $\hat{H}_{{\rm imp}}$,
\begin{equation}
	\hat{H}_{{\rm imp}}=-J_{\rm imp}{\displaystyle \sum_{x,y}}(\hat{a}_{x+1,y}^{\dagger}\hat{a}_{x,y}+\hat{a}_{x,y+1}^{\dagger}\hat{a}_{x,y}+{\rm H.c.}),
	\label{Himp}
\end{equation}
where $J_{\rm imp}$ denotes the impurity hopping parameter. The operators $\hat{a}_{x,y}$ ($\hat{a}^\dagger_{x,y}$) are the impurity's annihilation (creation) operators at site $(x,y)$. The impurity interacts with the host particles through the interaction Hamiltonian $\hat{H}_{{\rm U}}$,
\begin{equation}
	\hat{H}_{{\rm U}}=U{\displaystyle \sum_{x,y}}\hat{a}_{x,y}^{\dagger}\hat{a}_{x,y}\hat{b}_{x,y}^{\dagger}\hat{b}_{x,y},
    \label{HU}
\end{equation}
where $U>0$ represents the interaction strength. Due to the repulsive nature of the impurity-host interaction, the presence of a pinned quasihole can give rise to a bound state between the impurity and the corresponding topological excitation, as we will analyze below. Before addressing the many-body problem, it is useful to first review the bound states of a single particle confined in a two dimensional setting.

\paragraph{Bound states in 2D---}
A bound state occurs when a particle is confined to a finite region by an external potential, exhibiting a localized wave function and a binding energy lower than the potential at spatial infinity. It is well known that in one dimension (1D), any arbitrarily weak attractive potential produces at least one bound state~\cite{griffiths}. Similarly, at least one bound state is always guaranteed in 2D for an attractive potential [see Appendix]. For example, considering a circular potential well of radius $R$ and depth $V_{0}$, solving the time-independent Schr\"odinger equation in polar coordinates straightforwardly leads to the binding energy~[see Appendix]
\begin{equation}
	E_B^{2D}=\frac{\hbar^{2}z^{2}}{2m_{\rm imp}R^{2}}-V_0.
    \label{E2D}
\end{equation}
Here, $\hbar$ is the reduced Planck constant and $m_{\rm imp}$ is the particle mass. The dimensionless parameter $z$ is the solution of the transcendental equation $\mathcal{J}_{1}(z)\mathcal{K}_{0}(\sqrt{z_{0}^{2}-z^{2}})/[\mathcal{J}_{0}(z)\mathcal{K}_{1}(\sqrt{z_{0}^{2}-z^{2}})]\!=\!\sqrt{(z_{0}/z)^{2}-1}$ with $z_0\!=\!R\sqrt{2mV_0}/\hbar$, and $\mathcal{J}_{0,1} (\mathcal{K}_{0,1})$ denote the Bessel functions of the first (second) kind. Due to the behavior of this transcendental equation at small arguments, a 2D circular trap always possesses at least one bound state~[see Appendix].

\begin{figure}
	\centering\includegraphics[width=0.99\linewidth]{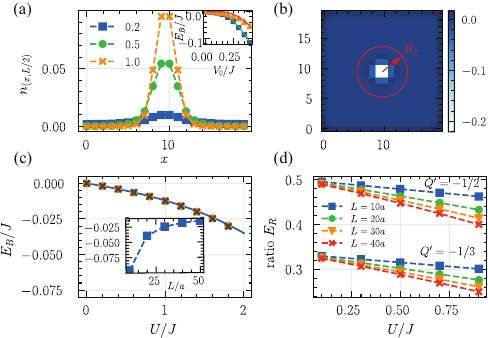}
	\caption{(a) The spatial density distribution of a single particle (along the center row) in finite system under periodic boundary condition with different trapping potentials $V_0/J=0.2,0.5$ and $1.0$. Inset: the binding energy $E_{2D}/J$ as a function of trap strength $V_0/J$ for different impurity hopping strength $J_{\rm imp}/J=0.6$ (squares) and $J_{\rm imp}/J=1$ (triangles). The solid lines represent the analytic results of Eq.(\ref{E2D}) after the substitution $m_{\rm imp}=\hbar^2/2J_{\rm imp}a^2$. The numerical results are  associated with the Hamiltonian $\hat{H}'_{\rm imp}$ with $R_1=2a$ and system size $20\times20$ under periodic boundary condition.  (b) The spatial density distribution in a Chern insulator with pinning potentials $V_{\rm pin}=2J$. (c) The binding energy as a function of $U/J$ in systems of size $L=20a$ under open boundary conditions. We plot the binding energy for a single particle subject to a pure trap of radius $R_1/a=1$ (solid line), a pure integer charge (dots), and the full density distribution of the CI ground state (crosses). Inset: binding energy as a function of system size $L$ at $U=2J_{\rm imp}$ in the case of using the integer charge. (d) The ratio of binding energies of the system with trapping potentials mimicking renormalized charge of value $Q'=-1/2, -1/3$ and that of the integer charge $Q=-1$. Despite finite size effect, the ratio of the binding energies approaches the charge values in the weak interaction limit.} 	
	\label{fig_MF}
\end{figure}

Such behavior also persists in a two-dimensional lattice. We consider a single impurity confined by a circular potential, described by the Hamiltonian $\hat{H}'_{{\rm imp}}=\hat{H}_{{\rm imp}}-V_{0} \sum_{x,y\in\mathcal{D}_1}\hat{a}_{x,y}^{\dagger}\hat{a}_{x,y}$, where $\mathcal{D}_1$ denotes the spatial region of the external potential, taken to be a disk of radius $R_1$ centered on the central plaquette of the lattice. Using $R_1=2a$ (with $a$ being the lattice constant) as an example, we show the impurity density distributions in Fig.~\ref{fig_MF}(a) for different confinement strengths $V_0$. As expected, weaker confinement $V_0$ leads to a more extended bound-state wave function and a smaller binding energy $E_B=E_{\rm gs}(V_0)-E_{\rm gs}(0)$, where $E_{\rm gs}(V_0)$ denotes the ground state energy of the Hamiltonian $\hat{H}'_{\rm imp}$ with $V_0$. This binding energy can also be described by the analytic result Eq.~(\ref{E2D}) in the continuum limit $m_{\rm imp}=\hbar^2/2J_{\rm imp}a^2$; see the inset of Fig.~\ref{fig_MF}(a). In both the weak and strong confinement limits, the binding energy exhibits an approximately linear dependence on $V_0$. This suggests a practical way to extract information about confining potentials, and in particular the effective fractional charge associated with a topological trapping medium (see below), by analyzing the binding energies.

\paragraph{Mean field analysis---}
We now consider the impurity-host interaction term in Eq.~(\ref{HU}) at the mean field level, i.e.\ $\hat{H}_{\rm U}\rightarrow\hat{H}_{\rm U}^{MF}=U \sum_{x,y}\langle\hat{n}_{x,y}\rangle\hat{a}_{x,y}^{\dagger}\hat{a}_{x,y}$. Here $\langle\hat{n}_{x,y}\rangle=n_{x,y}^{V_{h}}$ represents the density of the host particles associated with the ground state of the Hamiltonian $\hat{H}_{\rm host}$ at a given value of $V_h$. We denote this effective single impurity Hamiltonian as $\hat{H}''_{{\rm imp}}=\hat{H}_{{\rm imp}}+\sum_{x,y}Un_{x,y}^{V_{h}}\hat{a}_{x,y}^{\dagger}\hat{a}_{x,y}$.  We note that $Un_{x,y}^{V_{h}}$ plays a similar role to the external trap $-V_0$ in the Hamiltonian $\hat{H}'_{\rm imp}$ discussed in the previous paragraph.

To gain more insight into how topological excitations influence the impurity, we first investigate the distribution of an integer charge excitation in a CI, corresponding to the non-interacting fermionic case of Eq.\ (\ref{Hhost}).  Setting a flux density $\alpha=1/4$, and filling the lowest band, gives rise to a CI with Chern number $C\!=\!1$. It is characterized by a uniform density distribution in the incompressible bulk. Introducing a repulsive pinning potential $V_h$ creates a localized excitation carrying a negative charge $Q$, which is defined as: 
\begin{equation}
    Q=\sum_{x,y\in \mathcal{D}_2}(n_{x,y}^{V_{h}}-n_{x,y}^{V_{h}=0}),
    \label{Q}
\end{equation}
where $\mathcal{D}_{2}$ denotes a disk-shaped region of radius $R_{2}$ used to accumulate the density difference. For a typical value of $V_h/J=2$ and a sufficiently large radius $R_2\geq 4a$, we find a saturated charge value of $Q=-1$, consistent with the Chern number $C=1$; see Appendix. 
This negative charge distribution [Fig.~\ref{fig_MF}(b)] exhibits a similar shape to the pure trap with radius $R_1/a=1$ in the Hamiltonian $\hat{H}_{\mathrm{imp}}^{\prime }$. 
Using the charge distribution within $R_2=4a$ (to neglect the effect of edge excitations), as shown in Fig.~\ref{fig_MF}(c), we find good agreement between the binding energies associated with $\hat{H}_{\mathrm{imp}}^{\prime }$ (solid line) and $\hat{H}_{\mathrm{imp}}^{\prime \prime }$ (orange dots) after matching their integrated trap strengths via $-\sum_{x,y\in\mathcal{D}_1}V_0=-4V_0=UQ$. 
Furthermore, when using the full density distribution $Un_{x,y}^{V_{h}}$ (which includes the edge excitations), the binding energy remains in good agreement with the pure trap result, as indicated by the crosses in Fig.~\ref{fig_MF}(c). These results suggest that neither the detailed shape of the effective potential nor the presence of edge excitations affects the impurity bound state in the bulk.

The dependence of the binding energy on the external trap strength (related to $UQ$) provides a route to extract the charge value $Q$. To demonstrate this, we replace the integer charge $Q=-1$ by fractional values, e.g. $Q'=-1/2$ and $-1/3$, and calculate the binding energy ratio $E_R\equiv E_B(Q')/E_B(Q=1)$ as a function of $U$. As shown in Fig.~\ref{fig_MF}(d), the ratio approaches the charge value $|Q'|$ in the weak-interaction limit $U\!\rightarrow\!0$. In the weak-confinement regime, although a large system size is required to reach the saturated binding energy [see the inset of Fig.~\ref{fig_MF}(d)], the charge can still be reliably extracted from the binding-energy ratio in relatively small systems. This mean-field result provides a promising route for probing the many-body fractionalized excitations discussed below.

\begin{figure}
	\centering\includegraphics[width=0.98\linewidth]{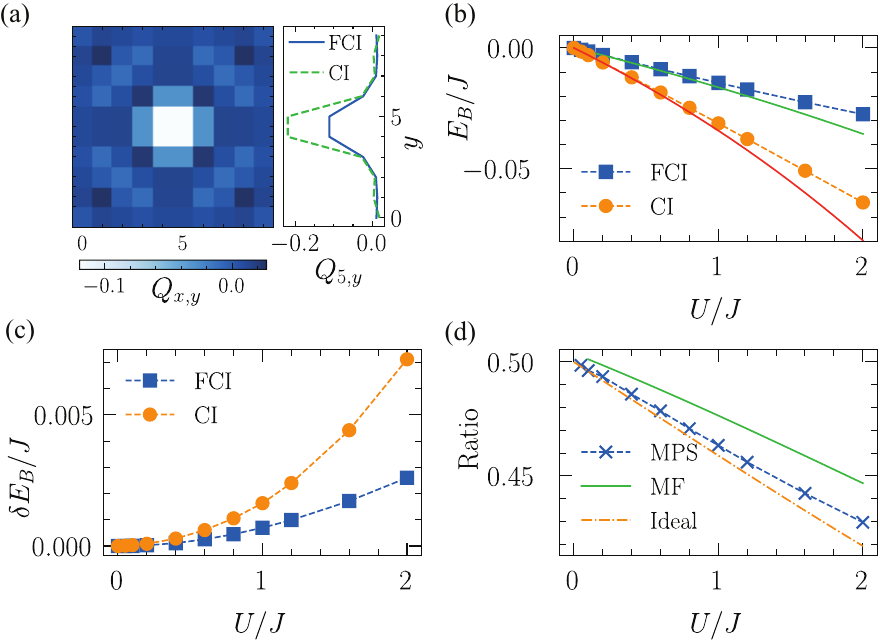}
	\caption{(a) Spatial distribution of the fractional charge $Q=-1/2$ in the $\nu=1/2$-FCI. Right: Charge density profiles along the central cut. The dashed line shows the corresponding integer-charge distribution in the CI. (b) Binding energy $E_B$ as a function of the impurity-host interaction strength $U$ for the CI (orange dots) and FCI (blue squares). The green and red solid curves denote the corresponding mean-field results obtained from $\hat{H}''_{\rm imp}$. (c) The binding energy difference $\delta E_B=E_B'-E_B$  as a function of interaction strength $U$.  (d) Ratio of binding energies $E_B(\nu=1/2)/E_B(\nu=1)$ as a function of $U/J$. The calculations are performed for $N=10$ particles in the $\nu=1/2$ FCI and $N=21$ free fermions in the CI on a $10\times10$ lattices with flux density $\alpha=1/4$ per plaquette. Blue crosses denote many-body results obtained using MPS; the green solid line shows mean-field results based on the host density distribution extracted from the MPS simulations; the orange dash-dotted line represents the single-particle analysis corresponding to the upper blue squares in Fig.~\ref{fig_MF}(d).} 	
	\label{fig_FCI}
\end{figure}

\paragraph{Many-body calculations---}
While the fractional values of $Q'$ considered above were introduced as illustrative examples on the level of a mean-field trapping potential, such charges naturally emerge as elementary excitations of FCIs. To demonstrate this, we consider hard-core bosons described by the host Hamiltonian $\hat{H}_{\rm host}$, with $N=10$ particles on a $10\times10$ lattice at flux density $\alpha=1/4$. In the absence of pinning potentials ($V_h=0$), the ground state realizes a Laughlin-type FCI at filling $\nu=1/2$~[see Appendix]. Introducing a pinning potential with strength comparable to the tunneling amplitude $J$, a localized fractional charge $Q=-1/2$ is generated in the bulk, as illustrated in Fig.~\ref{fig_FCI}(a). For given values of the pinning strength $V_h$ and impurity-host interaction $U$, we calculate the ground-state energy $E_{V_h}(U)$ of the full Hamiltonian $\hat{H}$ [Eq.~(\ref{H})] by optimizing an MPS approximation for the ground state using the DMRG algorithm \cite{1992White}. 

The binding energy $E_B$ is then defined as~\cite{Mostaan2026,2026Lu}
\begin{equation}
	E_{B}=\Delta E_{V_{h}}-\Delta E_{0},
	\label{EB}
\end{equation}
where $\Delta E_{V_{h}}\!=\!E_{V_{h}}(U)\!-\!E_{V_{h}}(0)$ represents the energy difference between a separated and bounded object formed by the fractional charge and the impurity. Since the fractional charge constitutes an excitation on top of the FCI ground state, we also subtracted the corresponding background contribution, $\Delta E_{0}=E_{0}(U)-E_{0}(0)$ in Eq.~\eqref{EB}.

 Figure~\ref{fig_FCI}(b) shows the binding energy as a function of the impurity-host interaction strength $U$. In the weak-coupling regime, the numerical results are in good agreement with the mean-field prediction. Similar behavior is obtained for free fermions described by Eqs.~(\ref{H}) and (\ref{Hhost}), corresponding to an impurity coupled to an integer charge excitation with $Q=-1$ in a CI.  At intermediate to strong impurity-host interactions $U$, the impurity's coupling to collective excitations of the medium (e.g.~magnetorotons) may become important~\cite{excitons_QH_dressing_1992}. To account for these effects, we introduce an alternative definition of the binding energy experienced by the impurity, 
\begin{equation}
	E'_B=\langle V_{h},U|\hat{H}_{{\rm imp}}+\hat{H}_{{\rm U}}|V_{h},U\rangle-\langle0,U|\hat{H}_{{\rm imp}}+\hat{H}_{{\rm U}}|0,U\rangle,
\end{equation}
where $|V_{h},U\rangle$ denotes the ground state of the full Hamiltonian $\hat{H}$ for given values of $V_h$ and $U$. 
While $E_B$ denotes the bare anyon-impurity binding energy, $E'_B$ is its effective binding energy within the many-body state, which additionally accounts for back-action effects. Their difference, $\delta E_B=E_B'-E_B$ [see Appendix], therefore provides a direct estimate of this back-action, capturing the dressing of the impurity through its coupling to the collective modes of the many-body background (e.g.~magnetorotons). As shown in Fig.~\ref{fig_FCI}(c), the difference $\delta E_B=E'_B-E_B$ increases with $U$, indicating the onset of collective excitations induced by the impurity-host interaction.
Nevertheless, $\delta E_B$ nearly vanishes in the weak-coupling limit ($U/J<0.5$), which suggests that the impurity predominantly binds  to a single quasihole in this regime, leaving the background essentially intact.

Since the binding energy is determined by the confinement strength, which is directly proportional to the excitation charge, the ratio $E_B(Q=-1/2)/E_B(Q=-1)$ provides a direct measure of the fractional charge in the weak-interaction limit.  This key feature is demonstrated in Fig.~\ref{fig_FCI}(d). Since adiabatic preparation of FCI states relies on a finite-size energy gap~\cite{Leonard2023,2024WangCan}, the clear fractional-charge signature already observed in a modest $10\times10$ system is encouraging, highlighting the feasibility of extracting such topological properties from impurity binding energies in quantum-engineered platforms.

\begin{figure}
	\centering\includegraphics[width=\linewidth]{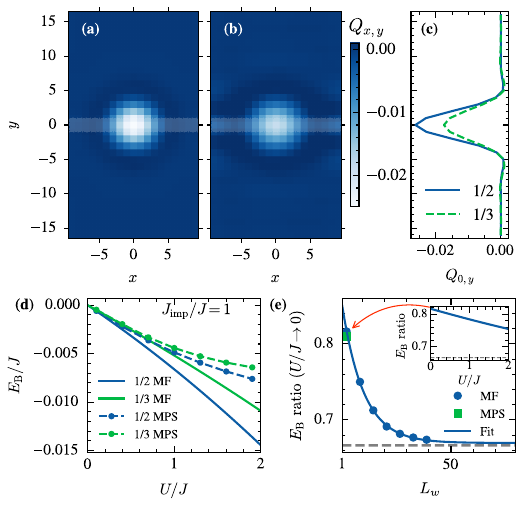}
	\caption{Spatial distribution of the fractional charge (a) $Q=-1/2$ and (b) $Q=-1/3$ associated with domain wall excitation using infinite MPS with circumference $L_x=19$. (c) Charge density profiles along the central cut. (d) Binding energy $E_B$ as a function of the impurity-host interaction strength $U$ for $\nu=1/2$ and $1/3$ within window size $L_w=3$. (e) Ratio of binding energies at weak interaction limit ($U/J=0.01$) as a function of the window size. Inset: Ratio of binding energies for window size $L_w=3$ as a function of $U$.} 	
	\label{fig_infinite}
\end{figure}

\paragraph{Domain wall excitation and infinite MPS---}
As an alternative approach, we also employ the window MPS approach \cite{Phien2012, Milsted2013, Zaletel2013}, which allows us to consider a localized fractional charge embedded in a FCI background on an infinite-cylinder geometry. On the cylinder, the FCI background takes the form of the Tao-Thouless states \cite{1983Tao} which develop crystalline order that decays with the cylinder circumference, on top of which the quasihole is realized as a domain wall excitation \cite{2005Bergholtz, 2005Seidel} that can be modeled as a window MPS \cite{palm2025fractional}. As shown in Figs.~\ref{fig_infinite}(a,b), fractional charges of $-1/2$ and $-1/3$ are obtained for hard-core bosons and fermions with nearest-neighbor interactions in Eq.\ (\ref{Hhost}), respectively; see the Appendix for technical details of the window MPS implementation. We observe similar charge distributions for the $-1/2$ and $-1/3$ quasiholes along the infinite $y$-direction [Fig.~\ref{fig_infinite}(c)], although the $-1/3$ quasihole exhibits more pronounced finite-size effects along the circumferential $x$-direction. 

Introducing a mobile impurity within a localized window of width $L_w$ [indicated by the shaded region in Figs.~\ref{fig_infinite}(a,b)] allows it to bind to the fractional charge. The resulting binding energies as functions of the impurity--host interaction strength $U$ are shown in Fig.~\ref{fig_infinite}(d). The MPS results (dashed dots) agree with the mean-field predictions (solid lines) in the weak interaction regime.
For weak impurity--host interactions $U$, we find that the ratio of the two binding energies, $E_B(-1/3)/E_B(-1/2)$, approaches the expected charge ratio of $2/3$ for sufficiently large window sizes $L_w$ [Fig.~\ref{fig_infinite}(e)]. Such finite-size effects can be attributed to the more extended charge distribution of the $-1/3$ quasihole along the $x$ direction, as shown in Fig.~\ref{fig_infinite}(b). We note that a similar window MPS approach was recently used for studying the formation of anyon molecules in the continuum ~\cite{2026Wang}.

\begin{figure}
	\centering\includegraphics[width=0.99\linewidth]{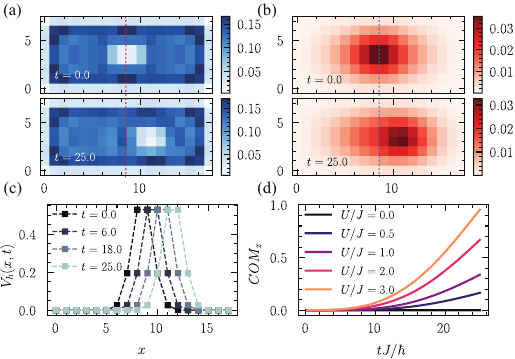}
	\caption{Spatial density distributions of (a) the host particles and (b) the impurity at evolution times $tJ/\hbar=0.0,12.0$ and $25.0$ (from top to bottom) for $U/J=3.0$. (c) Spatial profiles of the host pinning potentials along the row $y=3$  at different times. (d) Center-of-mass of the impurity as a function of evolution time for different interaction strengths $U$.} 	
	\label{fig_dyn}
\end{figure}

\paragraph{Dragging the anyon-impurity composite---}
While the evaluation of the binding energy allows for the extraction of fractional charges, the formation of an anyon-impurity bound state also offers a practical setting for manipulating anyonic excitations. For example, coherent transport of the anyon-impurity composite could enable measurements of anyonic braiding phases~\cite{2012Kapit,Palm2026}. Motivated by recent advances in manipulating anyons using antidots in electronic systems~\cite{2026Kim,2026Henzinger}, as well as the growing capabilities of combining optical lattices with optical tweezers~\cite{2024Braun}, we investigate the controlled transport of anyon-impurity composites through dynamical tuning of the pinning potential.

We begin with an anyon-impurity bound state and transport it by dynamically moving the pinning potential applied to the host particles. As an illustrative example, we consider an $8\times18$ lattice containing $N=14$ hard-core bosons. The local pinning potentials $V_h$ in $\hat{H}_{\rm host}$ is replaced by a Gaussian profile along the $x$-direction, i.e. $V_{h}(x,t)=V_{h}e^{-(x-x_{c}(t))^2/2\sigma^2}$ applied only to the rows $y=3,4$. Here $x_c$ denotes the center of the Gaussian potential. A localized fractional charge is generated around $x_c$ by setting $V_h/J=0.6$, as shown in Fig.~\ref{fig_dyn}(a). Starting from the ground state of the full Hamiltonian $\hat{H}$ at finite pinning strength, we simulate the real-time dynamics using an MPS approximation of the time-evolved state with the time-dependent variational principle (TDVP) algorithm \cite{Haegeman2016}.
During the evolution, the Gaussian pinning potential is translated along the $x$ direction by three lattice constants over a time interval $\tau$. The spatial profiles of the pinning potential at different times are shown in Fig.~\ref{fig_dyn}(c).

The resulting host-particle density distributions under the time-dependent pinning potential are shown in Fig.~\ref{fig_dyn}(a). As the fractional charge is transported along the $x$ direction, the impurity follows its motion and exhibits a corresponding displacement of its center of mass [Figs.~\ref{fig_dyn}(b,d)]. As expected, increasing the impurity--host interaction strength $U$ enhances the binding between the impurity and the fractional charge, thereby improving the transport efficiency. 
For $U/J=3$, the impurity center of mass is displaced by approximately one lattice constant after $\tau\simeq25\hbar/J$. More adiabatic transport can be achieved by slowing the dragging protocol to a timescale comparable to the inverse binding energy (e.g.\ $\simeq 34\hbar/J$ for $U/J=3$), at the cost of increased computational resources. Conversely, this direct link between the dragging speed and the binding energy suggests a practical route to extract the anyon-impurity binding energy dynamically, by monitoring the breakdown of adiabatic following of the impurity as the quasihole excitation is displaced~[see Appendix].

\paragraph{Conclusion---} We have investigated the formation and manipulation of impurity–anyon bound states in quantum-engineered fractional Chern insulators, where a local quasihole excitation and a mobile impurity can be finely controlled. Combining analytical arguments with large-scale numerical simulations, we have characterized the binding energy and stability of the composite object and shown that its binding energy provides a direct probe of the anyon's fractional charge. Furthermore, we demonstrated that the impurity–anyon bound state can be coherently transported by dynamically dragging the quasihole pinning potential, providing a practical route for controlled manipulation of fractionalized excitations~\cite{2026Kim,2026Henzinger}. 

This scheme is directly compatible with existing quantum simulators of the interacting Harper-Hofstadter model~\cite{2023Leonard,impertro2025strongly,2024WangCan}, and could equally be explored in the continuum~\cite{2020Clark,mukherjee2022crystallization,2024Lunt}. Combined with efficient schemes for preparing FCIs~\cite{Blatz2024bayesian,Palm2024,2025Wu,steinfadt2026dissipation,de2026adiabatic}, the local quasihole excitation can be created and dynamically steered using a programmable potential generated by a digital-micromirror device~\cite{2024Braun}, with its fractional charge inferred directly from \textit{in-situ} density measurements~\cite{Raciunas2018,Macaluso2020,Wang2022,Palm2024}. The mobile impurity can be realized by addressing a distinct internal (hyperfine) state of the same atomic species~\cite{2017Meinert,2025Dhar}, with the impurity-host interaction strength tuned via a Feshbach resonance~\cite{2008Bloch}. The anyon-impurity binding energy can in turn be extracted dynamically, by monitoring the breakdown of adiabatic following of the impurity as the quasihole excitation is displaced~[see Appendix], or through complementary techniques such as polaron spectroscopy~\cite{2026Massignan}. Our results establish a realistic framework for creating, detecting, and transporting anyon-impurity composites, paving the way toward experimentally accessible protocols for anyon braiding~\cite{Palm2026} in quantum-engineered platforms.

\textit{Acknowledgments---}
 We thank Ivan Amelio, Ignacio Cirac, Nader Mostaan, and Atac Imamoglu for discussions. We acknowledge support by the FRS-FNRS (Belgium), the ERC Starting Grant LATIS, the EOS project CHEQS, the Fondation ULB and the ANR PEPR Grant QUTISYM ANR-23-PETQ-0002. The numerical simulations were performed using ITensor~\cite{2022_itensor} (finite MPS calculations) and TensorKit \cite{MPSKit} (infinite MPS calculations).





\bibliography{bibliography,mybib}

\clearpage






\appendix

\begin{center}
{\large {\bf APPENDIX}}
\end{center}

\section{Binding energy from anyon-impurity drag}

Dragging the anyon-impurity bound state could be used to measure anyonic braiding phases in experiments~\cite{Palm2026}. However, monitoring this adiabatic motion can also be employed to estimate the anyon-impurity binding energy, as we now discuss. 

\begin{figure}[H]
	\centering\includegraphics[width=1.0\linewidth]{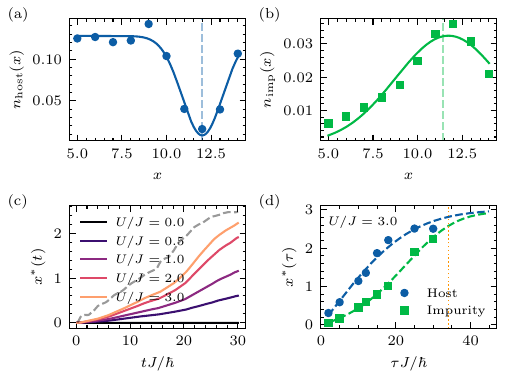}
	\caption{Spatial density distributions of (a) the host particles and (b) the impurity along the central row $y=4$ within the bulk for $U=3J$ and time $t=30\hbar/J$. The solid lines show the corresponding (inverted) Gaussian fits. The vertical dashed lines indicate the positions of the minimum and maximum, denoted by $x^*$. (c) Peak position of the impurity as a function of evolution time for different impurity--host interaction strengths $U$. The dashed line denotes the position of the host-particle density minimum, which is independent of $U$. (d) Positions of the host-particle density minimum (blue dots) and the impurity density maximum (green squares) as a function of total dragging time $\tau$ for fixed $U=3J$. Dashed lines represent the exponential polynomial fits. The convergence of these two occurs around $\tau = 40 \ \hbar/J$, which is close to the inverse binding energy of $\simeq 34 \ \hbar/J$ (orange dotted line).
    }
	\label{fig_peak}
\end{figure}

In Fig.~\ref{fig_peak}(a,b), we determine the maximum position of the impurity and the minimum position of the host particles by fitting the bulk density with (inverted) Gaussian functions. Since stronger impurity-host interaction $U$ leads to stronger binding, the impurity moves closer to the host-particle density minimum (i.e.~the quasihole position) as $U$  increases; see Fig.~\ref{fig_peak}(c). For fixed $U$, we track the positions of the host-particle density minimum and the impurity density maximum for different total evolution times $\tau$. As shown in Fig.~\ref{fig_peak}(d), these two start to converge when $\tau$ get closer to the inverse binding energy (indicated by the orange dashed line). These results suggest a practical route to measuring the anyon-impurity binding energy in quantum-engineered systems.

\section{Estimating the dressing by gapped collective modes}\label{App_estm}
Based on the definition of $E_B$ and $E'_B$ in the main text, the binding energy difference $\delta E_B=E_B'-E_B$ can be straightforwardly expressed as 
\begin{align}
    \delta E_B= &\langle V_{h,}U|\hat{H}_{{\rm host}}|V_{h},U\rangle-\langle0,U|\hat{H}_{{\rm host}}|0,U\rangle \nonumber \\ &
    -[\langle V_{h,}0|\hat{H}_{{\rm host}}|V_{h},0\rangle-\langle0,0|\hat{H}_{{\rm host}}|0,0\rangle].
\end{align}
Here $|V_{h},U\rangle$ is the ground state of the full Hamiltonian $\hat{H}$ for given values of $V_h$ and $U$. 
Since the subtrahend $[\langle V_{h,}0|\hat{H}_{{\rm host}}|V_{h},0\rangle-\langle0,0|\hat{H}_{{\rm host}}|0,0\rangle]$ represents the quasihole energy (for appropriate values of $V_h$), the quantity $\delta E_B$ captures the additional energy cost induced in the host system by the impurity–host interaction $U$. In particular, this contribution can be attributed to the back-action of the impurity onto the surrounding many-body state. In this sense, the difference $\delta E_B$ [Fig.~\ref{fig_FCI}(c)] provides a direct estimate of this back-action, and reflects the additional dressing of the impurity through its coupling to collective modes of the many-body background (e.g.~magnetoroton modes).

\section{Bound state in 2D}
\label{Appd_bound}

We consider a single particle confined in a two-dimensional (2D) circular box trap. The corresponding time-independent Schr\"odinger equation is
\begin{equation}\label{eq:schrodinger_eq}
	\left[-\frac{\hbar^2}{2M} \nabla^2 + V(r)\right] \Psi(r,\phi) = E \Psi(r,\phi),
\end{equation}
where $M$ is the mass of the particle, $\Psi$ is the wavefunction in polar coordinates $(r,\phi)$ and $V(r)$ is a radially symmetric potential given by
\begin{equation}
	V(r) = \begin{cases}
		-V_0 & \text{for } 0 \leq r \leq R \\
		0 & \text{for } r > R
	\end{cases}.
\end{equation}
Here $R$ is the radius of the circular box trap, and the Laplacian in polar coordinates reads
\begin{equation}
	\nabla^2 = \frac{1}{r} \frac{\partial}{\partial r} \left( r \frac{\partial}{\partial r} \right) + \frac{1}{r} \frac{\partial^2}{\partial \phi^2}.
\end{equation}

Using separation of variables $\Psi(r,\phi) = \psi(r) \chi(\phi)$ in \eqref{eq:schrodinger_eq}, the wavefunction splits into radial and azimuthal components,
\begin{equation}
	\frac{r}{\psi(r)} \frac{d}{dr} \left( r \frac{d\psi(r)}{dr}\right) + \frac{2Mr^2}{\hbar^2}\left( E - V(r) \right) = -\frac{1}{\chi(\phi)} \frac{d^2\chi(\phi)}{d\phi^2}.
\end{equation} 
Setting both the radial and azimuthal differential equations equal to a constant $c$, the azimuthal equation becomes
\begin{equation}
	\frac{d^2 \chi(\phi)}{d \phi^2} = -c \chi(\phi).
\end{equation}
Using the boundary conditions $\chi(0) = \chi(2\pi)$ and $\chi'(0) = \chi'(2\pi)$, we obtain $c= m^2$ with $m \in \mathbb{Z}$ an integer, and the azimuthal wavefunction $\chi(\phi) = e^{im\phi}$.

The radial differential equation can now be expressed as
\begin{align}
	& r^2 \frac{d^2\psi(r)}{dr^2} + r \frac{d\psi(r)}{dr} \nonumber \\ & \qquad + \left(\frac{2Mr^2}{\hbar^2}\left[ E - V(r) \right] - m^2\right) \psi(r) =0. \label{eq:radial_diff_eq}
\end{align}
Inside the circular trap $r \leq R$, the solution of Eq.~(\ref{eq:radial_diff_eq}) should remain finite as $r\rightarrow0$ and is found to be the Bessel function of the first kind,
\begin{equation}
	\psi_{\rm in}(r) = A \mathcal{J}_m(kr),
\end{equation}
with $k = \sqrt{2M(E + V_0)}/\hbar$, where we have assumed negative binding energy $<-V_0<E<0$. 

Outside the trap $r > R$, the solution must decay to zero at infinity, in which case the solution of Eq.~(\ref{eq:radial_diff_eq}) is the modified Bessel function of the second kind,
\begin{equation}
	\psi_{\rm out}(r) = B \mathcal{K}_m(\kappa r),
\end{equation}
with $\kappa = \sqrt{2M|E|}/\hbar$.

At the boundary $r = R$, the wave function and its first derivative must be continuous. This leads to the following trancendental equation
\begin{align}\label{eq:transcendental_equation}
	k\frac{\mathcal{J}'_m(ka)}{\mathcal{J}_m(ka)} = \kappa\frac{ \mathcal{K}'_m(\kappa a)}{\mathcal{K}_m(\kappa a)}.
\end{align}

\begin{figure}[H]
	\centering\includegraphics[width=0.9\linewidth]{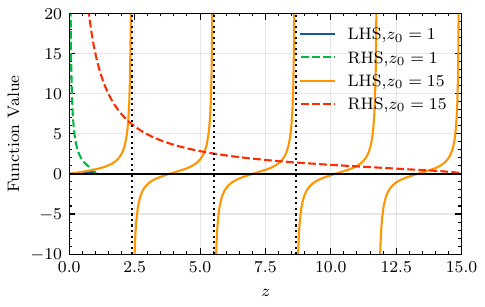}
	\caption{Left-hand side (LHS) and right-hand side (RHS) of the transcendental equation for two different values of $z_0$. Since the LHS approaches zero as $z\rightarrow0$, while the RHS vanishes at $z=z_0$, an arbitrarily weak confinement strength $V_0$, corresponding to an arbitrarily small $z_0$, guarantees at least one solution.} 	
	\label{fig_supp_bound}
\end{figure}

For the ground state, we set the angular momentum $m=0$. Using the Bessel identities $\mathcal{J}_0'(r) = -\mathcal{J}_1(z)$ and $\mathcal{K}_0'(z) = -\mathcal{K}_1(z)$, we obtain the trancendental equation,
\begin{equation}
	\frac{\mathcal{J}_{1}(z)\mathcal{K}_{0}\left(\sqrt{z_{0}^{2}-z^{2}}\right)}{\mathcal{J}_{0}(z)\mathcal{K}_{1}\left(\sqrt{z_{0}^{2}-z^{2}}\right)}=\sqrt{\left(\frac{z_{0}}{z}\right)^{2}-1},
	\label{eq_gs_tran}
\end{equation}
where $z_0 = \sqrt{2MV_0}R/\hbar$ and $z = kR$. The latter equation, together with the expression for $k$, gives rise to the binding energies
\begin{equation}
	E=\frac{\hbar^{2}z^{2}}{2MR^{2}}-V_0.
\end{equation}
Since the transcendental equation (\ref{eq_gs_tran}) always admits at least one solution $z$ [see Fig.~\ref{fig_supp_bound}], at least one bound state is guaranteed to exist, regardless of how shallow the confinement potential $V_0$ is. 

\section{Integer charge in a Chern insulator}
\begin{figure}
	\centering\includegraphics[width=0.95\linewidth]{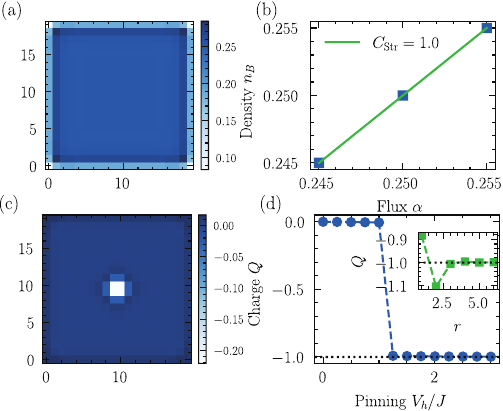} 
	\caption{(a) Spatial density distribution of a CI ground state. We consider a $20\times20$ lattice with uniform flux $\alpha=1/4$ and $N=96$ non-interacting fermions occupying the lowest band with Chern number $C=1$. (b) Bulk density, defined within a central disk of radius $3a$, as a function of flux $\alpha$. The slope yields the St\v{r}eda marker $C_{\rm Str}=1.0$. (c) Spatial distribution of the density difference between the CI ground states with pinning strengths $V_h/J=2.0$ and $V_h/J=0.0$. (d) Integrated charge within a central disk of radius $r=4a$ as a function of the pinning strength $V_h/J$. Inset: integrated charge as a function of disk radius $r$ for $V_h/J=2.0$.} 
	\label{fig_supp_CI}
\end{figure}

Consider non-interacting fermions in the Harper-Hofstadter lattice described by Eq.~(\ref{Hhost}) with $\hat{H}_{{\rm int}}^{{\rm B/F}}=0$. For a magnetic flux of $\alpha=1/4$ flux quantum per plaquette, filling the lowest band realizes a Chern insulator with Chern number $C=1$. This state is characterized by a uniform density distribution in the incompressible bulk. A representative spatial density distribution for a lattice with open boundary conditions is shown in Fig.~\ref{fig_supp_CI}(a). The topological nature can be further verified using the St\v{r}eda formula~\cite{1982Widom,1982Streda,1983Streda,2008Umucalilar,2020Repellin},
\begin{equation}
	C_{\text{Str}}=\frac{\partial n_{\text{B}}}{\partial\alpha}=\frac{\sigma_{\text{H}}}{\sigma_{0}},\label{eq_Streda}
\end{equation}
where $\sigma_0\!=\!1/2\pi$ is the conductivity quantum and $\sigma_{\text{H}}$ denotes Hall conductance. For the CI state with $C\!=\!1$ , the St\v{r}eda marker is expected to take the quantized value $C_{\text{Str}}\!=\!1$, as shown in Fig.~\ref{fig_supp_CI}(b).

To create a charge excitation, we apply a repulsive pinning potential $V_h$ at the central plaquette. The resulting spatial density distribution for $V_h/J=2$ is shown in Fig.~\ref{fig_supp_CI}(c). Integrating the density difference over a disk of radius $r$ centered on the pinning potential yields a saturated charge of $Q\simeq -1$ for $r\gtrsim 3a$, as shown in the inset of Fig.~\ref{fig_supp_CI}(d). The charge evaluated at $r=4a$ as a function of the pinning strength $V_h$ is plotted in Fig.~\ref{fig_supp_CI}(d). An abrupt jump to $Q=-1$ is observed once $V_h$ becomes comparable to the Harper--Hofstadter band gap. By contrast, the induced charge remains negligible for weak pinning potentials, reflecting the incompressibility of the insulating bulk.

\section{Fractional charge in an fractional Chern insulator}
\begin{figure}
	\centering\includegraphics[width=0.95\linewidth]{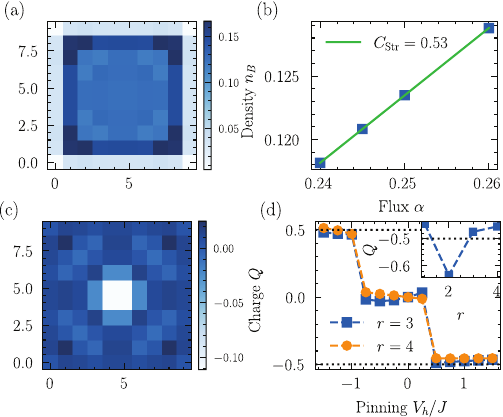} 
	\caption{(a) Spatial density distribution of the FCI ground state. The host system consists of $N=10$ hard-core bosons in a $10\times10$ lattice with a uniform flux $\alpha=1/4$. (b) Bulk density, defined within a central disk of radius $3a$, as a function of flux $\alpha$. The slope yields the St\v{r}eda marker $C_{\rm Str}=0.53\pm0.001$. (c) Spatial distribution of the density difference between the FCI ground states with pinning strengths $V_h/J=1.0$ and $V_h/J=0.0$. (d) Integrated charge within central disks of radii $r=3a$ and $4a$ as a function of the pinning strength $V_h/J$. Inset: integrated charge as a function of disk radius $r$ for $V_h/J=1.0$.} 
	\label{fig_supp_FCI}
\end{figure}

To realize the $\nu=1/2$ fractional Chern insulating state~\cite{Soerensen2005,Hafezi2007}, we consider hard-core bosons described by Eq.~(\ref{Hhost}) with $\hat{H}_{{\rm int}}^{{\rm B}}\!=\!\frac{U_{h}}{2} \sum_{x,y}\hat{n}_{x,y}(\hat{n}_{x,y}-1)$, where the on-site interaction strength is taken to the hard-core limit, $U_h\rightarrow\infty$. For $10$ particles in a $10\times10$ lattice with flux $\alpha=1/4$ and open boundary conditions, the spatial density distribution of the ground state is shown in Fig.~\ref{fig_supp_FCI}(a). Applying the St\v{r}eda formula within the bulk, we obtain $C_{\rm Str}\simeq0.53$, in good agreement with the expected many-body Chern number $C_{\rm MB}=1/2$.

Fractional charge excitations can also be created by applying on-site pinning potentials $V_h$ at the central plaquette. The resulting spatial charge distribution for $V_h/J=1$ is shown in Fig.~\ref{fig_supp_FCI}(c). Integrating the charge over a disk of radius $r$ centered on the pinning potential gives rise to a fractional value. We plot the charge values for $r/a=3$ and $4$ as a function of pinning strength $V_h$ in Fig.~\ref{fig_supp_FCI}(d). Fractional charges of $+1/2$ and $-1/2$ are obtained for negative and positive pinning potentials $V_h$, respectively. In the main text, we focus on the representative case $V_h/J=1$, which realizes a localized charge of $-1/2$. The repulsive impurity-host interaction ($U>0$) binds the impurity to the negative fractional charge. Analogous binding physics is expected for an attractive impurity--host interaction ($U<0$), which binds the impurity to the positive fractional charge.

\section{Domain-wall excitations in Tao-Thouless states}
Tao-Thouless (TT) states describe the degenerate ground states of strongly interacting quantum particles in the fractional quantum Hall (FQH) regime when the two-dimensional (2D) physical space is compactified along one direction into a thin-cylinder limit (or a thin torus, in the case of periodic boundary conditions) \cite{1983Tao, 2005Bergholtz, 2006Seidel}. This mapping reduces the 2D FQH problem to an effective quasi-one-dimensional (quasi-1D) problem, in which the ground states exhibit a charge-density-wave (CDW)-like crystalline pattern along the cylinder axis. In the strict 1D limit, the particles localize into fixed spatial orbitals, forming a CDW pattern that mimics the fractional filling of the underlying topological state; for the Laughlin $\nu = 1/2$ state, this pattern takes the form $\ldots 101010 \ldots$. TT states are adiabatically connected to Laughlin states: as the cylinder circumference increases, the CDW amplitude decays exponentially, and the crystalline density pattern melts into the uniform density characteristic of the 2D Laughlin state. Topological defects in the CDW pattern --- domain-wall excitations of the TT states --- carry fractional charge and are equivalent to the anyonic quasiparticle/quasihole excitations of the 2D state, realized here in the quasi-1D TT regime \cite{palm2025fractional}.

To realize this construction numerically, we obtain TT-equivalent representations of the bosonic $\nu = 1/2$ and fermionic $\nu = 1/3$ fractional Chern insulator (FCI) states on an infinite cylinder by variationally optimizing the ground state of the interacting Harper-Hofstadter model using $U(1)$-symmetric tensor-network methods that conserve total particle number. Specifically, we optimize an infinite MPS (iMPS) with unit cell $L = L_x L_y$ using the VUMPS algorithm \cite{2018Zauner-Stauber}, where $L_x$ is the cylinder circumference and $L_y = 2$ ($3$) for the hard-core boson $\nu = 1/2$ (fermion $\nu = 1/3$) state. The flux per plaquette is set to $\phi = 2\pi/L_x$, with particle density per site $n = 1/L$. For fermions, we additionally include a nearest-neighbor repulsion $V \sum_{\langle x,x' \rangle \langle y,y'\rangle} \hat{n}_{x,y} \hat{n}_{x',y'}$ to stabilize the topologically ordered FCI/TT ground state. Convergence was verified with a singular-value cutoff of $10^{-6}$.

We next construct a localized quasihole excitation using a finite-MPS ansatz, the window MPS, with a pinning potential fixing its position to a single site. A window MPS $X_n$ consists of $n$ iMPS unit cells with a fixed left environment given by one degenerate TT ground state, $\psi_L$, and a fixed right environment given by another, $\psi_R$, drawn from the same degenerate manifold. For the $\nu = 1/2$ state, for example, the two degenerate ground states correspond to the patterns $\ldots ABABAB\ldots$ ($\psi_L$) and $\ldots BABABA \ldots$ ($\psi_R$); pairing them within $X_n$ introduces a localized topological defect. We variationally optimize the ground state of $X_n$ using DMRG algorithm, yielding the fractionally charged anyonic excitations of the topologically ordered FCI/TT state. The fractional charge is localized by applying a pinning potential within the spatial manifold of $X_n$; we use a positive pinning potential $V_h = 5J$ to stabilize the quasihole. To get an idea for the charge of the quasihole we use $Q = \sum_{x,y} Q_{x,y}$ summing along the full infinite cylinder, here $Q_{x,y} = \left(\hat{n}_{x,y} - 1/L \right)$; we observe that as the circumference $L_x$ increases (and the CDW amplitude decreases) $Q \rightarrow \nu$ for Laughlin-like Fractional Quantum Hall states. In Fig.~\ref{fig_infinite}(a)-(b) we show the spatial distribution of the quasihole (domain wall excitation) for $\nu=1/2$ and $1/3$ FCI state with $L_x = 19$ (for which the CDW amplitude is vanishingly small and TT state approximates the 2D Fractional Quantum Hall state quite well). 

We then introduce a mobile impurity confined to the finite window $X_n$. By construction of the iMPS unit cells, the impurity is subject to periodic boundary conditions along the cylinder circumference ($x$ direction) and open boundary conditions along $y$; its range of longitudinal motion is set by the number of unit cells in $X_n$; $L_w = n L_y$. Switching on an onsite interaction between the impurity and the background host particles causes the pinned quasihole to act as an effective attractive potential, trapping the impurity and forming a quasihole-impurity bound state. The binding energy (and spatial extent) of this bound state depend on the strength of the impurity-particle interaction. Note that in Fig.~\ref{fig_infinite}, for comparison with $\nu=1/3$ state, we confined the impurity within window size $L_w=3$ for $\nu=1/2$ state by using an infinite positive potential on the last rung of the second iMPS unit cell used for creating the window MPS.

\section{An elongated fractional Chern insulator}
\begin{figure}[H]
	\centering\includegraphics[width=0.99\linewidth]{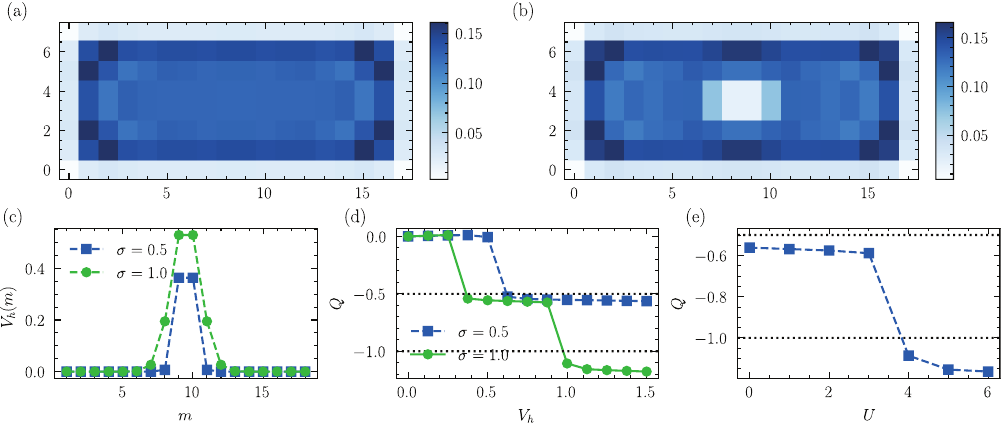} 
	\caption{Spatial density distributions of the FCI ground state with (a) vanishing pinning potential, $V_h=0$, and (b) finite pinning strength, $V_h/J=0.6$. The host system consists of $N=14$ hard-core bosons in an $18\times8$ lattice with uniform flux $\alpha=1/4$. (c) Gaussian profiles of the applied pinning potentials. (d) Integrated charge within the central rectangular region of size $6\times4$ as a function of pinning strength $V_h$ for $U=0$. (e) Integrated charge within the central rectangular region of size $6\times4$ as a function of impurity--host interaction strength $U$ for $V_h/J=0.6$ and $\sigma=1.0$.} 
	\label{fig_supp_drag}
\end{figure}

Here we investigate the ground-state properties of an elongated FCI relevant to the dragging dynamics. We consider $N=14$ hard-core bosons on an $8\times18$ Harper--Hofstadter lattice with flux $\alpha=1/4$, whose ground state realizes a $\nu=1/2$ FCI. To create fractional charges for the dragging protocol, we introduce Gaussian pinning potentials along the central rows (see the definition in the main text). As shown in Figs.~\ref{fig_supp_drag}(c,d), broader pinning potentials ($\sigma=1.0$) induce a fractional charge of approximately $-1/2$ for $0.25<V_h/J<1$ and charge $-1$ for $V_h/J\geq1$. To isolate a single fractional charge of $-1/2$, we fix $V_h/J=0.6$, for which the impurity--host interaction strength is restricted to $U/J\leq3.0$ [Fig.~\ref{fig_supp_drag}(e)].

\end{document}